\documentstyle[12pt]{article}
\newlength{\extraspace}
\newlength{\extraspaces}
\newcommand{\be}{\begin{equation}
\addtolength{\abovedisplayskip}{\extraspaces}
\addtolength{\belowdisplayskip}{\extraspaces}
\addtolength{\abovedisplayshortskip}{\extraspace}
\addtolength{\belowdisplayshortskip}{\extraspace}}
\newcommand{\ee}{\end{equation}}

\newcommand{\ba}{\begin{eqnarray}
\addtolength{\abovedisplayskip}{\extraspaces}
\addtolength{\belowdisplayskip}{\extraspaces}
\addtolength{\abovedisplayshortskip}{\extraspace}
\addtolength{\belowdisplayshortskip}{\extraspace}}
\newcommand{\ea}{\end{eqnarray}}

\newcommand{\nonu}{\nonumber \\[.5mm]}
\newcommand{\A}{&\!\!\!}

\newcommand{\newsection}[1]{
\vspace{7mm} \pagebreak[3] \addtocounter{section}{1}
\setcounter{subsection}{0} \setcounter{footnote}{0}
\begin{center}
{\large {\bf \thesection. #1}}
\end{center}
\nopagebreak
\medskip
\nopagebreak \hspace{3mm}}

\begin{document}

\begin{center}
{\bf Brane World black holes in Teleparallel Theory Equivalent to
General Relativity and their Killing vectors, Energy, Momentum and
Angular-Momentum}\footnote{Mathematics Department, Faculty of
Science, Ain Shams University, Cairo, Egypt.}
\end{center}
\centerline{ Gamal G.L. Nashed}

\bigskip

\centerline{\it Centre for Theoretical Physics, The British
University in Egypt,
 El-Sherouk City,}
\centerline{{\it Misr - Ismalia Desert Road, Postal No. 11837,
P.O. Box 43, Egypt.}}
\bigskip

 \centerline{ e-mail:gamal.nasshed@bue.edu.eg}

\hspace{2cm}
\\
\\
\\
\\
\\
\\
\\
\\

The energy-momentum tensor, which is coordinate independent, is
used to calculate energy, momentum and angular-momentum of two
different tetrad fields. Although, the two tetrad fields reproduce
the same space-time their energies are different. Therefore, a
regularized expression of the gravitational energy-momentum tensor
of the teleparallel equivalent of general relativity, (TEGR), is
used to make the energies of the two tetrad fields equal.  The
definition of the gravitational energy-momentum is used to
investigate the energy within the external event horizon. The
components of  angular-momentum associated with these space-times
are calculated. In spite that we use a static space-times, we get
a non-zero component of angular-momentum! Therefore, we  derive
the killing vectors associated with these space-times using the
definition of the Lie derivative of a second rank tensor in the
framework of the TEGR to make the picture more clear.\vspace{.2cm}\\
\hspace*{.1cm}{\it Keywords: Teleparallel equivalent of general
relativity, Brane world black holes, Gravitational energy-momentum
tensor, Regularized expression of the gravitational energy-momentum.}\vspace{0.01cm}\\\\
\hspace*{0.1cm}{\it PACS numbers: 0440, 0450, 0455, 0490.}
\\
\\
\\
\\
\\
\\
\\

\begin{center}
\newsection{\bf Introduction}
\end{center}

Quantum mechanics and general relativity are two very successful
and well validated theories within their own domains. The problem
is that there is no way to unify them into a single consistent
theory. One of the most promising models of unification is string
theory. Two classes of strings are the closed and open strings.
Gravity is described by closed strings and matter is described by
open strings. In non-pertubative string theory there exist
extended objects known as D-branes. These are surfaces where the
open strings must start and finish. This provides an alternative
to Kaluza-Klein \cite{KK} approach, where matter penetrates the
extra dimensions, leading to strong constraints from collider
physics.

A model that captures some of the essential features of the
dimensional reduction of 11-dimensional supergravity proposed by
Ho$\tilde{r}$ava and Witten \cite{HW} is introduced \cite{RR}. The
gravitational field on the brane is defined by the modified
Einstein equations given by Shiromizu, Maeda and Sasaki \cite{SMS}
from 5-dimensional gravity using the Gauss and Codazzi equations
\cite{GC}. \be G_{\mu \nu}=-\Lambda_4\delta_{\mu \nu}-K^2_4 T_{\mu
\nu}-K^4_5\Pi_{\mu \nu}-E_{\mu \nu}, \quad
\Lambda_4=\frac{\kappa^2_5}{2}\left(\Lambda_5+\frac{\varrho^2
\kappa^2_5}{6}\right),\ee where $\Lambda_4$ is the 4-dimensional
cosmological constant expressed in terms of the 5-dimensional
cosmological constant $\Lambda_5$, $G_{\mu \nu}$ is the Einstein
4-dimensional tensor,   $\varrho$ is the brane tension,
$\kappa^2_4=8\pi G_N=\frac{\varrho\kappa^2_5}{6\pi}$ is the
4-dimensional  gravitational constant, $G_N$ is the Newton's
constant of gravity, $T_{\mu \nu}$ is the stress energy tensor of
matter confined on the brane, $\Pi_{\mu \nu}$ is a tensor
quadratic in $T_{\mu \nu}$ obtained from the 5-dimensional metric
across the brane \be
2{\Pi_\mu}^\nu={T_\mu}^\beta{T_\beta}^\nu-T{T_\mu}^\nu-{\delta_\mu}^\nu\left(T_{\rho
\epsilon}T^{\rho \epsilon}-\frac{T^2}{2}\right),\ee where $T$ is
the trace of the stress energy tensor of matter and $E_{\mu \nu}$
is the electric part of the 5-dimensional Weyle tensor projected
onto the brane. In proper 5-dimensional coordinate $E_{\mu
\nu}={\delta_\mu}^M{\delta_\nu}^NC_{MNKL}n^Kn^L$ where $M,N
\cdots$ are 5-dimensional indices and $n^M$ is the unit normal to
the brane \cite{BGCF}. Here we are going to select a class of
spherically symmetric black holes without specifying $E_{\mu
\nu}$.

Among  various attempts to overcome the problems of quantization
and the existence of singular solution in Einstein's general
relativity (GR), gauge theories of gravity are of special
attractive, as they based on the concept of gauge symmetry which
has been very successful in the foundation of other fundamental
interactions. The importance of the poincar$\acute{e}$ symmetry in
particle physics leads one to consider the poincar$\acute{e}$
gauge theory (PGT) as a natural framework for description of the
gravitational phenomena \cite{Kt}$\sim$\cite{HMM}.  Basic
gravitational variables in PGT are the tetrad field ${e^a}_\mu$
and the Lorentz connection ${A^{ab}}_\mu$, which are associated to
the translation and Lorentz subgroups of the poincar$\acute{e}$
group, respectively. These gauge fields are coupled to the
energy-momentum and spin of matter fields, and their field
strengths are geometrically identified with the torsion and the
curvature. The space-time of the PGT turns out to be
Riemann-Cartan space $U_4$, equipped with metric and linear,
metric compatible connection.

General geometric arena of PGT, the Riemann-Cartan space $U_4$,
may be a priori restricted by imposing certain conditions on the
curvature and the torsion. Thus, Einstein's GR is defined in
Riemann space $V_4$, which is obtained from $U_4$ by the
requirement of vanishing torsion. Another interesting limit of PGT
is the {\it teleparallel or Weitzenb$\ddot{o}$ck} geometry $T_4$.
The vanishing of the curvature means that parallel transport is
path independent. The teleparallel geometry is, in sense,
complementary to Riemannian: curvature vanishes, and torsion
remains to characterize the parallel transport. Of particular
importance for the physical interpretation of the teleparallel
geometry the fact that there is a one-parameter family of
teleparallel Lagrangians which is empirically equivalent to GR
\cite{HNV,HS9,Nj}. For the parameter value $B=1/2$ the Lagrangian
of the theory coincides, modulo a four-divergence, with the
Einstein-Hilbert Lagrangian, and defines (TEGR).

The search for a consistent expression for the gravitating energy
and angular-momentum of a self-gravitating distribution of matter
is undoubtedly a long-standing problem in GR. It is believed that
the energy of the gravitational field is not localizable, i.e.,
defined in a finite region of the space. The gravitational field
does not possess the proper definition of an energy momentum
tensor and one usually defines some energy-momentum and
angular-momentum as Bergmann \cite{BT} or Landau-Lifschitz
\cite{LL} which are pseudo-tensors and depend on the second
derivative of the metric tensor. These quantities can be annulled
by an adequate transformation of coordinate. They \cite{BT,LL}
justify the results as being consistent with Einstein's principle
of equivalence. In this principle, it can be always find a small
region of space-time, where it prevails Minkowski space-time.  In
such a space-time, energy of the gravitational field is null.
Therefore, it is only possible to define the energy of the
gravitational filed in whole space-time region and not only in a
small region. The Einstein's GR can also be reformulated in the
context of teleparallel geometry \cite{PP}$\sim$\cite{AGP1}. In
this geometrical setting the dynamical field quantities correspond
to orthonormal tetrad field ${e^a}_\mu$\footnote{space-time
indices $\mu, \ \ \nu, \cdots$ and SO(3,1) indices a, b $\cdots$
run from 0 to 3. Time and space indices are indicated to $\mu=0,
i$, and $a=(0), (i)$.} (a, $\mu$ are SO(3,1) and space-time
indices, respectively). The teleparallel geometry is a suitable
framework to address the notions of energy, momentum and
angular-momentum of any space-time that admits a $3+1$ foliation
\cite{MR}. Therefore, we consider the TEGR.

 Hamiltonian formulation of TEGR,
in Schwinger's time gauge \cite{Sj}, has been established
\cite{Mj}. This formulation is an alternative teleparallel
geometric description to Einstein GR theory. An essential feature
of this Hamiltonian formulation is that one can defines energy of
the gravitational field by means of an adequate interpretation of
the Hamiltonian constraint. Several configurations of the
gravitational energy were investigated with success \cite{Mj,RC}.
TEGR can be understood as a gauge theory for the translation group
\cite{AGP1}. In this approach, the gravitational interaction is
described by a force similar to the Lorentz force equation of
electrodynamics, with torsion playing the role of force.  Regge
and Teitelboim \cite{RT} obtained a Hamiltonian formalism for GR
that is manifestly invariant under Ponicar$\acute{e}$
transformations at infinity by introducing ten new pairs of
canonical variables, which yield ten surface integrals to the
total Hamiltonian. The subsequent analysis given by York
\cite{Yoj} showed that a proper definition for the gravitational
angular-momentum requires a suitable asymptotic behavior of the
spatial components of the Ricci tensor. A careful analysis of the
exact form of the boundary conditions needed to define the energy,
momentum and angular-momentum of the gravitational field has been
carried out by Beig and  $\acute{o}$ Murchadha \cite{BM} and by
Szabados \cite{SLB} who found the necessary conditions that yield
a finite value for the above quantities. In these analysis the
poincar$\acute{e}$ algebra are realized at the spacelike infinity.
These are transformations of the Cartesian coordinates in the
asymptotic region of the space-time.

The Hamiltonian formulation of {\it an arbitrary teleparallel
theories }using Schwinger's time gauge is established
\cite{MS2,SM2}. In this formulation it is shown that the TEGR is
the {\it only viable consistent teleparallel theory of gravity}.
Maluf and Rocha \cite{MD} established a theory in which {\it
Schwinger's time gauge has not been incorporated in the geometry
of absolute parallelism}. In this formulation, the definition of
the gravitational angular-momentum arises by suitably interpreting
the integral form of the constraint equation $\Gamma^{ab}=0$. This
definition has been successfully applied to the gravitational
field of a thin, slowly rotating mass shell \cite{MUFR} and for
the three-dimensional BTZ black hole \cite{MPR}. Shanxian  et al.
\cite{SJr} have calculated the energy of general 4-Dim. stationary
axisymmetric spacetime in the teleparallel geometry.

Definitions for the gravitational energy in the context of the
TEGR have already been proposed in the literature. An expression
for the gravitational energy arises from the surface term of the
total Hamiltonian is given in Ref. \cite{Nj8}. In Ref.
\cite{BV2001}, a similar quantity is suggested. Both expressions
are equivalent to the integral form of the total divergences of
the Hamiltonian density developed in Ref. \cite{MD}. The three
expressions yield the same value for the total energy of the
gravitational field. However, since these three expressions
contain the lapse function in the integrand, non of them is
suitable to the calculation of the irreducible mass of the Kerr
black hole because the lapse function vanishes on the external
event horizon of the black hole \cite{Mj}. The energy expressions
of References  \cite{Nj8}, \cite{BV2001} are not to be applied to
a finite surface integration; rather they yield the total energy
of the space-time \cite{Mj}.

The Localization of gravitational energy-momentum remains an
important problem in GR. Using the standard methods many famous
researchers each found their own expression. None of these
expressions is covariant, they are all reference frame dependent
(referred to as pseudotensors). This feature can be understood in
terms of the equivalence principle: gravity cannot be detected at
a point, so it cannot have a point-wise defined energy-momentum
density. Now there is another way to address this difficulty. The
new idea is the quasi-local energy-momentum is associated with a
closed 2 surface surrounding a region. A good quasi-local approach
is in terms of the Hamiltonian \cite{NHC}. Then the Hamiltonian
boundary term determines the quasi-local quantities. In fact this
approach includes all the traditional pseudotensors
\cite{NHC,CN2}. They are each generated by a superpotential which
can serve as special type of Hamiltonian boundary term. A good
energy-momentum expression for gravitating systems should satisfy
a variety of requirements, including giving the standard values
for the total quantities for asymptotically flat space and
reducing to the material energy-momentum in proper limit. No
entirely expression has yet been identified. One of the most
restrictive requirements is positivity. A general positivity proof
is very difficult \cite{SNC}. For more details of the topic of
quasi-local approach a very nice review article is given in
\cite{SL}.

The definition of $P^a$ is invariant under global $SO(3,1)$
transformations. It has been argued elsewhere \cite{Mj5} that it
makes sense to have a dependence of $P^a$ on the frame. The
energy-momentum in classical theories of particles and fields does
not depend on the frame, and it has been asserted that such
dependence is a natural property of the gravitational
energy-momentum. The total energy of a relativistic body, depends
on the frame. It is assumed that a set of tetrads fields is
adapted to an observer in the space-time determined by the metric
tensor $g_{\mu \nu}$.

We investigate the irreducible mass $M_{irr}$ of the brane world
black holes. It is the total mass of the black hole at the final
stage of Penrose's process of energy extraction, considering that
the maximum possible energy is extracted. It is also related to
the energy contained within the external event horizon $E(r_+)$ of
the black hole (the surface of the constant radius $r=r_+$ defines
the external event horizon). Every expression for local or
quasi-local gravitational energy must necessary yield the value of
$E(r_+)$ in close agreement with $2M_{irr}$, since we know
beforehand the value of the latter as a function of the initial
angular-momentum of the black hole \cite{CD}. The evolution of
$2M_{irr}$ is a crucial test for any expression for the
gravitational energy. $E(r_+)$ has been obtained by means of
different energy expressions in Ref. \cite{BG}. The gravitational
energy used in this article is the only one that yields a
satisfactory value for $E(r_+)$ and that arises in the framework
of the Hamiltonian formulation of the gravitational field.

It is the aim of the present paper to find asymptotically flat
solutions with spherical symmetry  in the TEGR   for the
gravitational such that $\Lambda_4=T_{\mu \nu}=0$. In this case we
can treat Eq. (1) as conventional Einstein equations with an
effective stress energy tensor $\hat{T}_{\mu \nu}$. Using the
energy-momentum tensor \cite{Mj}, we calculate the energy,
momentum, angular-momentum associated with these solutions. We
also derive the killing vectors related to these solutions to
discuss the different results of energy, momentum,
angular-momentum.

 The paper is organizing as follows. In \S 2, we briefly review the
 TEGR theory, the energy-momentum tensor and the angular-momentum.  The two different
 tetrad fields with two unknown functions are
  studied in \S 2. In \S 3, we use the regularized expression for the
 gravitational energy-momentum to recalculate the energy. In \S 4,
 we derive explicitly the killing vectors related to these two tetrad
 fields, using the definition of the Lie derivative of a second rank tensor
in the framework of the TEGR,
 to discuss the different results we obtained for the energy and angular-momentum.
   Final section is devoted to discussion and conclusion.
\newsection{The  TEGR for gravitation, energy, momentum, angular-momentum}

In a space-time with absolute parallelism the parallel vector
fields ${e_a}^\mu$ define the nonsymmetric affine connection \be
{\Gamma^\lambda}_{\mu \nu} \stackrel{\rm def.}{=} {e_a}^\lambda
{e^a}_{\mu, \nu}, \ee where $e_{a \mu, \ \nu}=\partial_\nu e_{a
\mu}$\footnote{space-time indices $\mu, \ \ \nu, \cdots$ and
SO(3,1) indices a, b $\cdots$ run from 0 to 3. Time and space
indices are indicated to $\mu=0, i$, and $a=(0), (i)$.}. The
curvature tensor defined by ${\Gamma^\lambda}_{\mu \nu}$, given by
Eq. (3), is identically vanishing. The metric tensor $g_{\mu \nu}$
 is defined by
 \be g_{\mu \nu} \stackrel{\rm def.}{=}  \eta_{a b} {e^a}_\mu {e^b}_\nu, \ee
with $\eta_{a b}=(-1,+1,+1,+1)$ is the metric of Minkowski
space-time.

  The Lagrangian density for the gravitational field in the TEGR,
  in the presence of matter fields, is given by\footnote{Throughout this paper we use the
relativistic units$\;$ , $c=G=1$ and $\kappa=8\pi$.} \cite{Mj} \be
{\cal L}_G  =  e L_G =- \displaystyle {e \over 16\pi}  \left(
\displaystyle {T^{abc}T_{abc} \over 4}+\displaystyle
{T^{abc}T_{bac} \over 2}-T^aT_a
  \right)-L_m= - \displaystyle {e \over 16\pi} {\Sigma}^{abc}T_{abc}-L_m,\ee
where $e=det({e^a}_\mu)$. The tensor ${\Sigma}^{abc}$ is defined
by \be {\Sigma}^{abc} \stackrel {\rm def.}{=} \displaystyle{1
\over 4}\left(T^{abc}+T^{bac}-T^{cab}\right)+\displaystyle{1 \over
2}\left(\eta^{ac}T^b-\eta^{ab}T^c\right).\ee $T^{abc}$ and $T^a$
are the torsion tensor and the basic vector field  defined by \be
{T^a}_{\mu \nu} \stackrel {\rm def.}{=}
{e^a}_\lambda{T^\lambda}_{\mu
\nu}=\partial_\mu{e^a}_\nu-\partial_\nu{e^a}_\mu, \qquad \qquad
{T^a}_{b c} \stackrel {\rm def.}{=}  {e_b}^\mu {e_c}^\nu
{T^a}_{\mu \nu} \qquad \qquad T^a \stackrel {\rm
def.}{=}{{T^b}_b}^a.\ee The quadratic combination
$\Sigma^{abc}T^{abc}$ is proportional to the scalar curvature
$R(e)$, except for a total divergence term \cite{Mj}. $L_m$
represents the Lagrangian density for matter fields.

The gravitational  field equations for the system described by
${\it L_G}$ are the following
 \be  e_{a \lambda}e_{b \mu}\partial_\nu\left(e{\Sigma}^{b \lambda \nu}\right)-e\left(
 {{\Sigma}^{b \nu}}_a T_{b \nu \mu}-\displaystyle{1 \over 4}e_{a \mu}
 T_{bcd}{\Sigma}^{bcd}\right)= \displaystyle{1 \over 2}{\kappa} eT_{a
 \mu},\ee
where \[ \displaystyle{ \delta L_m \over \delta e^{a \mu}} \equiv
e T_{a \mu}.\] It  is possible to prove by explicit calculations
that the left hand side of the symmetric field equations (8) is
exactly given by \cite{Mj}
 \[\displaystyle{e \over 2} \left[R_{a
\mu}(e)-\displaystyle{1 \over 2}e_{a \mu}R(e) \right]. \] The
axial-vector part of the torsion tensor $a_\mu$ is defined by \be
a_\mu \stackrel{\rm def.}{=} {1 \over 6} \epsilon_{\mu \nu \rho
\sigma} T^{\nu \rho \sigma}={1 \over 3} \epsilon_{\mu \nu \rho
\sigma} \gamma^{\nu \rho \sigma}, \qquad where \qquad
\epsilon_{\mu \nu \rho \sigma} \stackrel{\rm def.}{=} \sqrt{-g}
\delta_{\mu \nu \rho \sigma}, \ee and $\delta_{\mu \nu \rho
\sigma}$ being completely antisymmetric and normalized as
$\delta_{0123}=-1$.

In the context of Einstein's GR, rotational phenomena is certainly
not a completely understood issue. The prominent manifestation of
a purely relativistic rotation effect is the dragging of inertial
frames. If the angular-momentum of the gravitational field of
isolated system has a  meaningful notion, then it is reasonable to
expect the latter to be somehow related to the rotational motion
of the physical sources.

The angular-momentum of the gravitational field has been addressed
in the literature by means of different approaches. The oldest
approach is based on pseudotensors \cite{BT,LL}, out of which
angular-momentum superpotentials are constructed. An alternative
approach assumes the existence of certain Killing vector fields
that allow the construction of conserved integral quantities
\cite{Ka}. Finally, the gravitational angular-momentum can also be
considered in the context of Poincar$\acute{e}$ gauge theories of
gravity \cite{HS8}, either in the Lagrangian or in the Hamiltonian
formulation. In the latter case it is required that the generators
of spatial rotations at infinity have a well defined functional
derivatives. From this requirement a certain surface integral
arises, whose value is interpreted as the gravitational
angular-momentum.

The Hamiltonian formulation of TEGR is obtained by establishing
the phase space variables. The Lagrangian density does not contain
the time derivative of the tetrad component $e_{a0}$. Therefore,
this quantity will arise as a Lagrange multiplier \cite{Dp}. The
momentum canonically conjugated to $e_{ai}$ is given by
$\Pi^{ai}=\delta L/\delta \dot{e}_{ai}$. The Hamiltonian
formulation is obtained by rewriting the Lagrangian density in the
form $L=p \ \dot{q}-H$, in terms of $e_{ai}, \Pi^{ai}$ and the
Lagrange  multipliers. The Legendre transformation can be
successfully carried out and the final form of the Hamiltonian
density has the form \cite{MR} \be H=e_{a0}C^a+\alpha_{ik}
\Gamma^{ik}+\beta_k\Gamma^k,\ee plus a surface term. Here
$\alpha_{ik}$ and $\beta_k$ are Lagrange multipliers that are
identified as \be \alpha_{ik}={1 \over 2} (T_{i0k}+T_{k0i}) \qquad
and \qquad \beta_k=T_{00k},\ee and $C^a$, $\Gamma^{ik}$ and
$\Gamma^k$ are first class constraints. The Poisson brackets
between any two field quantities $F$ and $G$ is given by \be \{
F,G \}=\int d^3x \left( \displaystyle{\delta F \over \delta
e_{ai}(x)} \displaystyle{\delta G \over \delta
\Pi^{ai}(x)}-\displaystyle{\delta F \over \delta
\Pi^{ai}(x)}\displaystyle{\delta G \over \delta e_{ai}(x)}
\right).\ee We recall that the Poisson brackets
$\left\{\Gamma^{ij}(x),\Gamma^{kl}(x)\right\}$ reproduce the
angular-momentum algebra \cite{Mj}.

 The constraint $C^a$ is
written as $C^a=-\partial_i \Pi^{ai}+h^a$, where $h^a$ is an
intricate expression of the field variables. The integral form of
the constraint equation $C^a=0$ motivates the definition of the
gravitational energy-momentum $P^a$ four-vector \cite{Mj} \be
P^a=-\int_V d^3 x
\partial_i \Pi^{ai},\ee where $V$ is an arbitrary volume of the
three-dimensional space. In the configuration space we have \ba
\Pi^{ai} \A =\A -4\kappa \sqrt{-g} \Sigma^{a0i} \quad with \quad
\partial_\nu(\sqrt{-g}\Sigma^{a \lambda \nu})=\displaystyle{1
\over 4 \kappa}\sqrt{-g}{e^a}_\mu (t^{\lambda \mu}+T^{\lambda
\mu}) \quad where \nonu
\A \A   t^{\lambda \mu}=\kappa \left(4\Sigma^{bc
\lambda}{T_{bc}}^\mu-g^{\lambda \mu} \Sigma^{bcd}T_{bcd}
\right).\ea

The emergence of total divergences in the form of scalar or vector
densities is possible in the framework of theories constructed out
of the torsion tensor. Metric theories of gravity do not share
this feature. By making $\lambda=0$ in Eq. (14) and identifying
$\Pi^{ai}$ in the left side of the latter, the integral form of
Eq. (14) is written as \be P^a=\int_V d^3 x \sqrt{-g}
{e^a}_\mu\left(t^{0 \mu}+T^{0 \mu} \right).\ee Eq. (15) suggests
that $P^a$ is now understood as the gravitational  energy-momentum
\cite{Mj}. The spatial component $P^{(i)}$ form a total
three-momentum, while temporal component $P^{(0)}$ is the total
energy \cite{LL}.

It is possible to rewrite the Hamiltonian density of Eq. (10) in
the equivalent form \cite{MUFR} \be H=e_{a0}C^a+\displaystyle{1
\over 2}\lambda_{ab}\Gamma^{ab}, \qquad with \qquad
\lambda_{ab}=-\lambda_{ba}, \ee are the Lagrangian multipliers
that are identified as $\lambda_{ik}=\alpha_{ik}$ and
$\lambda_{0k}=-\lambda_{k0}=\beta_k$.  The constraints
$\Gamma^{ab} = -\Gamma^{ba}$ \cite{MR} embodies both constraints
$\Gamma^{ik}$ and $\Gamma^k$ by means of the relation \be
\Gamma^{ik}={e_a}^i {e_b}^k \Gamma^{ab}, \qquad and \qquad
\Gamma^k \equiv \Gamma^{0k}={e_a}^0 {e_b}^k \Gamma^{ab}.\ee The
constraint $\Gamma^{ab}$ can be reads as \be
\Gamma^{ab}=M^{ab}+4\kappa\sqrt{-g}{e_{(0)}}^0
\left(\Sigma^{a(0)b}-\Sigma^{b(0)a}\right).\ee

In similarity to the definition of $P^a$, the integral form of the
constraint equation $\Gamma^{ab}=0$ motivates the new definition
of the space-time angular-momentum. The equation $\Gamma^{ab}=0$
implies \be M^{ab}=-4\kappa\sqrt{-g}{e_{c}}^0
\left(\Sigma^{acb}-\Sigma^{bca}\right),\ee Maluf et al. \cite{Mj,
MUFR} defined \be L^{ab} =\int_V d^3x {e_\mu}^a {e_\nu }^b M^{\mu
\nu }, \ee  as the 4-angular-momentum of the gravitational field
for an arbitrary volume V of the three-dimensional space. In
Einstein-Cartan type theories there also appear constraints that
satisfy the Poisson bracket as given by Eq. (12). However, such
constraints arise in the form $\Pi^{[ij]}=0$, and so a definition
similar to Eq. (20), i.e., interpreting the constraint equation as
an equation for the angular-momentum of the field, {\it is not
possible}. Definition (20) is three-dimensional integral. The
quantities $P^a$ and $L^{ab}$ are separately invariant under
general coordinate transformations of the three-dimensional space
and under time reparametrizations, which is an expected feature
since these definitions arise in the Hamiltonian formulation of
the theory. Moreover, these quantities transform covariantly under
global $SO(3,1)$ transformations \cite{MUFR}.

We will consider two simple configuration of  tetrad fields and
discuss their physical interpretation as reference frames. The
first one in quasi-orthogonal coordinate system  can be written as
\cite{Rh} \ba {{e_{(0)}}^0} \A= \A A, \quad {e_\alpha}^0 = C x^a,
\quad {e_{(0)}}^\alpha = D x^\alpha \nonu
{e_i}^\alpha \A= \A \delta_a^\alpha B + F x^a x^\alpha +
\epsilon_{a \alpha \beta} S x^\beta,
 \ea where {\it A}, {\it C},
{\it D}, {\it B}, {\it F}, and {\it S} are unknown functions of
${\it r}$. It can be shown that the functions $D$ and $F$ can be
eliminated by coordinate transformations \cite{HS7,SNH}, i.e., by
making use of freedom to redefine $t$ and $r$, leaving the tetrad
field (21) having four unknown functions in the quasi-orthogonal
coordinates. Thus the tetrad field (21) without the functions $D$
and $F$ and also without the two functions $C$ and $S$ will be
used in the following discussion  for the calculations of energy,
momentum and angular-momentum but in the spherical coordinate.
Therefore, the tetrad field (21) can be written in the spherical
coordinates without the functions $D$, $F$, $C$ and $S$ as
\cite{SNH,SA} \be \left({{e_1}_a}^{ \mu} \right) = \left( \matrix{
\frac{1}{A} &0 & 0 & 0 \vspace{3mm} \cr 0 & \frac{B}{\sqrt{r}}
\sin\theta \cos\phi & \frac{\cos\theta \cos\phi}{r}
 & -\frac{\sin\phi} {r \sin\theta} \vspace{3mm} \cr
0 & \frac{B}{\sqrt{r}} \sin\theta \sin\phi & \frac{\cos\theta
\sin\phi}{r}
 & \frac{\cos\phi} {r \sin\theta} \vspace{3mm} \cr
0 & \frac{B}{\sqrt{r}} \cos\theta & -\frac{\sin\theta}{r} & 0 \cr
} \right). \ee

The other configuration of tetrad field that has a simple
interpretation as a reference frame can has the form \be
\left({{e_2}_a}^{ \mu} \right) = \left( \matrix{ \frac{1}{A} &0 &
0 & 0 \vspace{3mm} \cr 0 & \frac{B}{\sqrt{r}} & 0
 & 0 \vspace{3mm} \cr
0 & 0 & \frac{1}{r}
 & 0 \vspace{3mm} \cr
0 & 0 & 0&\frac{1}{r \sin\theta}  \cr } \right). \ee

The space-time associated with the two tetrad fields (22) and (23)
is the same and has the form \be ds^2=-A^2 dt^2+\frac{r}{B^2}
dr^2+r^2(d\theta^2+\sin^2\theta d\phi^2),\ee and the non-vanishing
components of the effective stress energy-momentum tensor
associated with the space-time given by Eq. (24) are
\[{\hat{T_0}^0}=\hat{\rho}=\frac{1-2BB'}{\kappa_4^2r^2}, \qquad \qquad
{\hat{T_1}^1}=\hat{\rho}_{rad}=\frac{rA-B^2(A-2rA')}{\kappa_4^2
r^3A},\]
\[{\hat{T_2}^2}={\hat{T_3}^3}=\hat{\rho}_{tang}=\frac{B\left(AB-2rAB'-rBA'-2r^2BA''-
2r^2A'B'\right)}{2\kappa_4^2 r^3A}.\]

 The two tetrad fields
satisfy the field equations (8). Now we are going to calculate the
energy, momentum and angular-momentum  associated with the two
tetrad fields (22) and (23). For asymptotically flat space-times
$P^0$ yields the ADM energy \cite{ADM}. In the context of tetrad
theories of gravity, asymptotically flat space-times may be
characterized by the asymptotic boundary condition \be e_{a \mu}
\cong \eta_{a \mu} + \displaystyle{1 \over 2} h_{a \mu}(1/r),\ee
and by the condition $\partial_\mu {e^a}_\mu=O(1/r^2)$ in the
asymptotic limit $r \rightarrow \infty$. An important property of
tetrad fields that satisfy Eq. (25) is that in the flat space-time
limit one has ${e^a}_\mu(t,x,y,z)={\delta^a}_\mu$, and therefore
the torsion tensor ${T^a}_{\mu \nu}=0$.

We  apply Eq. (13) to the tetrad field (22) to calculate
 the energy content. Calculations are performed in the spherical coordinate.
  Eqs. (22) and (23)
 assumed that the reference space is determined by a set of tetrad
 fields ${e^a}_\mu$ for the flat space-time such that the
 condition ${T^a}_{\mu \nu}=0$ is satisfied. Using Eq. (7) in Eq.
 (22), the non-vanishing components of the torsion tensor  are
given by \ba {T^{(0)}}_{01} \A=\A \frac{A'}{A}, \qquad \qquad
{T^{(2)}}_{12}=\frac{(\sqrt{r}-B)}{rB}={T^{(3)}}_{13},\ea and the
non-vanishing component of the tensor $T^{(a)}$ is given by \be
T^{(1)}=\frac{B \left(rBA'-2A\{\sqrt{r}-B\} \right)}{r^2A}. \ee
The axial vector associated with Eq. (22) is vanishing identically
due to the fact that the tetrad field of Eq. (22) has a spherical
symmetry \cite{Rh}.

Now we are going to apply Eq. (6) to the tetrad field (22) using
Eqs. (26) and (27) to calculate
 the energy content. We perform the calculations in the spherical
 coordinate. The only required component of ${\Sigma}^{\mu \nu
 \lambda}$ is
 \be
{\Sigma}^{(0) 0 1}=-\frac{\sin\theta \{r-\sqrt{r}B\}}{4\pi }.\ee
Further substituting Eq. (28) in (13) we obtain \be
P^{(0)}=E=-\oint_{S \rightarrow \infty}
 dS_k \Pi^{(0) k}=-\displaystyle {1  \over 4 \pi} \oint_{S \rightarrow \infty}
 dS_k  e
{\Sigma}^{(0) 0 k}= {\{r-\sqrt{r}B\}}.\ee Let us apply expression
(13) to the evaluation of the irreducible mass by fixing $V$ to be
the volume within the $r=r_+$ surface where $r_+$ is the external
horizon, i.e., $B=0$. Therefore, \be P^{(0)}=E=-\int_{S_i} dS_i
\Pi^{(0)i}=-\int_S d\theta d\phi \Pi^{(0)1}(r,\theta,\phi),\ee
where the surface $S$ is determined by the condition $r=r_+$. The
expression of $\Pi^{(0)1}$ will be obtained by considering Eq.
(14) using Eq. (6) and Eq. (7). The expression of
$\Pi^{(0)1}(r,\theta,\phi)$ for the tetrad (22)
 reads \be \Pi^{(0)1}=\frac{\sin\theta \{r-\sqrt{r}B\}}{4\pi
}, \ee integrate Eq. (31) on the surface of constant radius
$r=r_+$ where $r_+$ is the external horizon of the  black hole. On
this surface the second term of Eq. (31) vanishes. Therefore, on
the surface $r=r_+$ we get \be P^{(0)}=E=r_+,\ee a result that is
obtained before \cite{Mj,SL}.

 Now let us continue to calculate the
 momentum and angular-momentum associated with the first tetrad
field given by Eq. (22).
 Using Eq. (14) in (22) we get \be \Pi^{(1) 1}=0.\ee Substitute Eq. (33) in Eq. (13) we
get \be P^{(1)} =\int_VdV
\partial_{1}(\Pi^{(1)1}(r,\theta,\phi))=
\int_S dS_1 \Pi^{(1)1}(r,\theta,\phi) = 0 .\ee  By the same method
we obtain \be \Pi^{(2) 1} =  0,   \qquad P^{(2)}= 0 ,\qquad
\Pi^{(3) 1} = 0, \qquad P^{(3)}= 0 . \ee The results of Eqs. (34)
and (35) are expected results since the space-time given by Eq.
(22) is a spherical symmetric  static space-time. Therefore, the
spatial momentum associated with any static solution is
identically vanishing \cite{MTW}.

We have used Eqs. (19) and (6) in Eq. (20) to calculate the
components of the angular-momentum. Finally we get \ba
M^{(1)(2)}\A=\A {e^{(1)}}_a {e^{(2)}}_b M^{ab}=M^{(1)(3)}=
{e^{(1)}}_a {e^{(3)}}_b M^{ab}=M^{(2)(3)}= {e^{(2)}}_a {e^{(3)}}_b
M^{ab}=0\nonu
M^{(0)(1)}\A=\A {e^{(0)}}_a {e^{(1)}}_b M^{ab}= -\frac{ \sqrt{r}
A^2 \sin^2\theta \cos\phi{\ \{r-\sqrt{r}B\}}}{4\pi B}, \nonu
M^{(0)(2)} \A=\A  {e^{(0)}}_a {e^{(2)}}_bM^{ab}=-\frac{\sqrt{r}A^2
\sin^2\theta \sin\phi{\ \{r-\sqrt{r}B\}}}{4\pi B},\nonu
 M^{(0)(3)} \A =\A {e^{(0)}}_a {e^{(3)}}_bM^{ab}=-\frac{\sqrt{r}A^2 \sin\theta \cos\theta{\
\{r-\sqrt{r}B\}}}{4\pi B}.\ea Using Eq. (36) in (20) we get
 \ba L^{(0)(1)} \A =\A {\int_0^\pi}{\int_0^{2\pi}}{\int_{0}^\infty}
 d\theta d\phi dr M^{(0)(1)}(r,\theta,\phi) = 0 ,\nonu
 \A \A by \ \ \ the \ \ \  same \ \ \  method \ \ \  we \ \ \  can \ \ \
 get \nonu
 L^{(0)(2)} \A=\A L^{(0)(3)} =
 L^{(1)(2)}=L^{(1)(3)} = L^{(2)(3)}= 0.\ea It is of interest to note that
the vanishing of $L^{(0)(1)}$, $L^{(0)(2)}$ is due to the
appearance of terms like  $\sin\phi$ and $\cos\phi$ while the
vanishing of $L^{(0)(3)}$ is due to the
 appearance of term like $\sin\theta \cos \theta$.

 Repeating the same calculations to the second tetrad given by Eq. (23)
 and write the necessary components of the torsion tensor and the vector
field $T^{(a)}$, we get \be {T^{(0)}}_{01}=\frac{A'}{A},\quad
{T^{(2)}}_{12}={T^{(3)}}_{13}=\frac{-1}{r},\quad  \quad
{T^{(3)}}_{23}=- \cot \theta,\ee and the only non-vanishing
components of the tensor $T^{(a)}$ are given by  \be
T^{(1)}=\frac{B^2\{2A+rA'\}}{r^2A}, \qquad T^{(2)}=\frac{\cot
\theta}{r^2}.\ee

The only required component of ${\Sigma}^{\mu \nu
 \lambda}$ needed to calculate the energy  is
 \be
{\Sigma}^{(0) 0 1}=\frac{ \sqrt{r}B \sin\theta}{4\pi}.\ee Further
substituting Eq. (40) into Eq. (13) we obtain \be
P^{(0)}=E=-\oint_{S \rightarrow \infty}
 dS_k \Pi^{(0) k}=-\displaystyle {1  \over 4 \pi} \oint_{S \rightarrow \infty}
 dS_k  e
{\Sigma}^{(0) 0 k}={\sqrt{r}B}.\ee By the same method used for the
first tetrad given by Eq. (22) we find that the  momentum and
angular-momentum associated with the second tetrad field given by
Eq. (23) have the form \ba \Pi^{(1) 1}\A =\A 0, \qquad P^{(1)}
=\int_VdV
\partial_{1}(\Pi^{(1)1}(r,\theta,\phi))=
\int_S dS_1 \Pi^{(1)1}(r,\theta,\phi) = 0, \nonu
  \Pi^{(2) 1} \A=\A 0,
\quad P^{(2)}= 0 , \quad \Pi^{(3) 1} = 0, \quad   P^{(3)}= 0, \ea

\ba M^{(1)(2)}\A=\A {e^{(1)}}_a {e^{(2)}}_b M^{ab}=M^{(1)(3)}=
{e^{(1)}}_a {e^{(3)}}_b M^{ab}=M^{(2)(3)}= {e^{(2)}}_a {e^{(3)}}_b
M^{ab}=M^{(0)(3)}= {e^{(0)}}_a {e^{(3)}}_b M^{ab}=0\nonu
 M^{(0)(1)}\A=\A {e^{(0)}}_a {e^{(1)}}_b M^{ab}=\frac{r A^2 \sin\theta }{4\pi}, \qquad \qquad
M^{(0)(2)}= {e^{(0)}}_a {e^{(2)}}_b M^{ab}=\frac{r^{3/2} A^2
\cos\theta }{8\pi B }.\ea Using Eq. (43) in (20) we get
 \ba L^{(0)(1)} \A=\A {\int_0^\pi}{\int_0^{2\pi}}{\int_{0}^\infty}
 d\theta d\phi dr M^{(0)(1)}(r,\theta,\phi) = -{\int_{0}^\infty}\frac{r^{3/2}A^2}{B} dr \neq 0
  ,\qquad by \ the
 \ same \ method \ we \ can \ obtain \nonu
 L^{(0)(2)} \A= \A L^{(0)(3)} =
 L^{(1)(2)}=L^{(1)(3)} = L^{(2)(3)}= 0.\ea It is of interest to note that
the vanishing of  $L^{(0)(2)}$ is due to the appearance of terms
like   $\cos\theta$. Here we obtain the component of the angular
momentum $L^{(0)(1)} \neq 0$ contradict what is well know that the
angular momentum of a spherically symmetric stationary spacetime
is vanishing \cite{MTW}.

 Let us gave a specific value for the
two unknown functions $A$ and $B$ to have the form \be
A=\displaystyle\sqrt{1-\frac{2M}{r}}, \qquad and
 \qquad B=\displaystyle\sqrt{\frac{(1-\frac{3M}{2r})}{(1-\frac{2M}{r})
 (1-\frac{\lambda_0}{r})}}.\ee
This  is a solution obtained before by Casadio et al. \cite{CFM}
in search for new brane world black holes and by Germani et al.
\cite{GM} as a possible external metric of a homogeneous star on
the brane. From Eq. (4) using the tetrad field (22) and Eq. (45)
we get the associated metric in  the form \be
ds^2=-\left(1-\frac{2M}{r}\right)
dt^2+\frac{(1-\frac{3M}{2r})}{(1-\frac{2M}{r})(1-\frac{\lambda_0}{r})}dr^2+r^2(d\theta^2
+\sin^2\theta d \phi^2).\ee The Schwarzschild metric is restored
in the special case $\lambda_0=\frac{3M}{2}$. In the case when
$\lambda_0>2M$ the metric of Eq. (46) describes a symmetric
traversable wormhole  \cite{BK}. If we apply Eq. (13) to the
tetrad given by Eq. (22) using (45) we get the energy to has the
form  \be P^{(0)} \cong \int_{r\rightarrow \infty} d\theta d\phi
\displaystyle{1 \over 16\pi}\sin \theta
\left(M+2\lambda_0\right)=\frac{M}{4}+\frac{\lambda_0}{2},\ee and
when $\lambda_0=\frac{3M}{2}$ we get
 \be P^{(0)} \cong M,\ee which is the ADM \cite{MTW}. The energy
 associated with the second tetrad given by Eq. (23) using Eq. (45) is given by
\be P^{(0)} \cong -r+\frac{M}{4}+\frac{\lambda_0}{2}!\ee which is
different from the energy given by Eq. (47).

As we see from Eqs. (47) and (49) that the energy associated with
the two tetrad field given be Eqs. (22) and (23) are different in
spite that they reproduce the same space-time as given by Eq.
(46). Definition of energy as given by Eq. (13) depends mainly on
the definition of torsion and the components of the torsion of
both tetrad fields given by Eqs. (26) and (38) which are
different. The flatness condition given by Eq. (25) of both tetrad
fields are satisfied when $A=1$ and $B=1$. The components of the
torsion associated with the two tetrad fields (22) and (23) are
now have the following form after using Eq. (45). For the tetrad
(22) we have
 \be {T^{(0)}}_{01}=\frac{M}{r(r-2M)}, \quad
 {T^{(2)}}_{12}=
 \frac{\sqrt{r(r-2M)(r-\lambda_0)(4r-6M)}-2(r-2M)(r-\lambda_0)}
{2r(r-2M)(r-\lambda_0)}={T^{(3)}}_{13},\ee and for the tetrad (23)
we have  \be {T^{(0)}}_{01}=\frac{M}{r{(r-2M)}},\quad
{T^{(2)}}_{12}=-\frac{1}{r}={T^{(3)}}_{13}, \qquad \qquad
{T^{(3)}}_{23}=- \cot \theta.\ee
 The space-time (46) is a flat space-time when $M=0$ and
$\lambda_0=0$ in this case the components of the torsion tensor of
the first tetrad are vanishing identically, $ {T^{(a)}}_{\mu
\nu}=0$ satisfying the flatness condition given by Eq. (25). The
components of torsion given by Eq. (51) of the second tetrad when
$M=0$ and $\lambda_0=0$ do not vanishing identically contradict
the flatness condition given by Eq. (25). Therefore, in this case
we are going to use the regularized expression for the
gravitational energy-momentum.
\newpage
\newsection{Regularized expression for the gravitational energy-momentum
  and localization of energy}

An important property of the tetrad fields that satisfy the
condition of Eq. (25) is that in the flat space-time limit
${e^a}_\mu(t,x,y,z)={\delta^a}_\mu$, and therefore the torsion
${T^\lambda}_{\mu \nu}=0$.  Hence for the flat space-time it is
normally to consider a set of tetrad fields such that
${T^\lambda}_{\mu \nu}=0$ {\it in any coordinate system}. However,
in general an arbitrary set of tetrad fields that yields the
metric tensor for the asymptotically flat space-time does not
satisfy the asymptotic condition given by (25). Moreover for such
tetrad fields the torsion ${T^\lambda}_{\mu \nu} \neq 0$ for the
flat space-time \cite{MVR}. It might be argued, therefore, that
the expression for the gravitational energy-momentum (13) is
restricted to particular class of tetrad fields, namely, to the
class of frames such that ${T^\lambda}_{\mu \nu}=0$ if ${E_a}^\mu$
represents the flat space-time tetrad field \cite{MVR}. To explain
this, let us calculate the flat space-time of the tetrad field of
Eq. (23) using (46) which is given by \be \left({{E_2}_a}^\mu
\right) =\left(\matrix {1&0 &0 &0 \vspace{3mm} \cr 0&1 &0& 0
\vspace{3mm} \cr 0& 0&\frac{1}{r}&0 \vspace{3mm} \cr
0&0&0&\frac{1}{r\sin\theta} \cr } \right). \ee Expression (52)
yields the following non-vanishing torsion components: \be
{T^{(2)}}_{12}=-\frac{1}{r}={T^{(3)}}_{13}, \qquad \qquad
{T^{(3)}}_{23}=- \cot \theta.\ee The tetrad field (52) when
written in the Cartesian coordinate will have the form \be
\left({{E_2}_a}^\mu(t,x,y,z) \right) =\left(\matrix {1&0 &0 &0
\vspace{3mm} \cr 0& \frac{x}{r} & \frac{y}{r} & \frac{z}{r}
\vspace{3mm} \cr 0 & \frac{xz}{r\sqrt{x^2+y^2}} &
\frac{yz}{r\sqrt{x^2+y^2}}& -\frac{\sqrt{x^2+y^2}}{r} \vspace{3mm}
\cr 0&-\frac{y}{\sqrt{x^2+y^2}}&\frac{x}{\sqrt{x^2+y^2}}& 0 \cr }
\right). \ee In view of the geometric structure of (54), we see
that, Eq. (23) does not display the asymptotic behavior required
by Eq. (25). Moreover, in general the tetrad field (54) is adapted
to accelerated observers \cite{MR,Mj,MVR}. To explain this, let us
consider a boost in the x-direction of Eq. (54).  We find \be
\left({{E_2}_a}^\mu(t,x,y,z) \right) =\left(\matrix {\gamma&v
\gamma &0 &0 \vspace{3mm} \cr \frac{v \gamma x}{r} & \frac{\gamma
x}{r} &\frac{y}{r} &\frac{z}{r} \vspace{3mm} \cr \frac{v \gamma
xz}{r\sqrt{x^2+y^2}}&\frac{\gamma xz}{r\sqrt{x^2+y^2}} &\frac{
yz}{r\sqrt{x^2+y^2}}&\frac{-\sqrt{x^2+y^2}}{r} \vspace{3mm} \cr
\frac{- v \gamma y}{\sqrt{x^2+y^2}}&\frac{-\gamma
y}{\sqrt{x^2+y^2}}&\frac{x}{\sqrt{x^2+y^2}}& 0 \cr } \right), \ee
where $v$ is the speed of the observer and
$\gamma=\frac{1}{\sqrt{1-v^2}}$. For a static object in a
space-time whose four-velocity is given by $u^\mu=(1,0,0,0)$ we
may compute its frame components $u^a={e^a}_\mu u^\mu=(\gamma,
\frac{v \gamma x}{r},\frac{v \gamma xz}{r\sqrt{x^2+y^2}},\frac{- v
\gamma y}{\sqrt{x^2+y^2}})$. It can be shown that along an
observer's trajectory whose velocity is determined by $u^a$  the
quantities \be {\phi_{(j)}}^{(k)}=u^i\left({{E_2}^{(k)}}_m
\partial_i {{E_2}_{(j)}}^m\right),\ee constructed out from (55) are
non vanishing. This fact indicates that along the observer's path
the spatial axis ${{E_2}_{(a)}}^\mu$ rotate \cite{MR,MVR}. In
spite of the above problems discussed for the tetrad field of Eq.
(23) it yields a satisfactory value for the total gravitational
energy-momentum, as we will discussed.

 In Eq. (13) it is implicitly assumed that the reference space is determined
 by a set of tetrad fields ${e^a}_\mu$ for flat space-time such
 that the condition ${T^a}_{\mu \nu}=0$ is satisfied. However, in
 general there exist flat space-time tetrad fields for which ${T^a}_{\mu \nu} \neq
 0$. In this case Eq. (13) may be generalized \cite{MR,MVR} by
 adding a suitable reference space subtraction term, exactly like
 in the Brown-York formalism \cite{BHS,YB}.

 We will denote ${T^a}_{\mu \nu}(E)=\partial_\mu {E^a}_\nu-\partial_\nu
 {E^a}_\mu$ and $\Pi^{a j}(E)$ as the expression of $\Pi^{a j}$
 constructed out of the flat tetrad ${E^a}_\mu$. {\it The
 regularized form of the gravitational energy-momentum $P^a$ is
 defined by} \cite{MR,MVR}
 \be P^a=-\int_{V} d^3x \partial_k \left[ \Pi^{a k}(e)-\Pi^{a k}(E)
 \right].\ee This condition guarantees that the energy-momentum of
 the flat space-time always vanishes. The reference space-time is
 determined by tetrad fields ${E^a}_\mu$, obtained from
 ${e^a}_\mu$ by requiring the vanishing of the physical parameters
 like mass, angular-momentum, etc. Assuming that the space-time is
 asymptotically flat then Eq. (57) can have the form \cite{MR,MVR}

\be P^a=-\oint_{S\rightarrow \infty} dS_k \left[ \Pi^{a
k}(e)-\Pi^{a k}(E) \right],\ee where the surface $S$ is
established at spacelike infinity. Eq. (58) transforms as a vector
under the global SO(3,1) group \cite{Mj}. Now we are in a position
to proof that the tetrad field (23) yields a satisfactory value
for the total gravitational energy-momentum.

We will integrate Eq. (58) over a surface of constant radius
$x^1=r$ and require $r\rightarrow \infty$. Therefore, the index
$k$ in (58) takes the value $k=1$. We need to calculate the
quantity
\[\Sigma^{(0) 01}={e^{(0)}}_0\Sigma^{0 01}=\displaystyle{1 \over
2}{e^{(0)}}_0(T^{001}-g^{00}T^{1}).\] Evaluate  the above equation
we find \be \Pi^{(0)1}(e)=\displaystyle{-1 \over 4\pi}e\Sigma^{(0)
01}=-\frac{1} {4\pi}\frac{\sin
\theta{\sqrt{r(r-2M)(2r-2\lambda_0)}}}{{\sqrt{2r-3M}}}
 \cong
\frac{-\sin \theta\left(4r-M-2\lambda_0\right)}{16\pi},\ee and the
expression of $\Pi^{(0)1}(E)$ is obtained by just making $M=0$ and
$\lambda_0=0$ in Eq.(59), it is given by \be
\Pi^{(0)1}(E)=\displaystyle{-1 \over 4\pi}r\sin(\theta).\ee Thus
the gravitational energy of the tetrad field of Eq. (23) is given
by \be P^0 \cong \int_{r\rightarrow \infty} d\theta d\phi
\displaystyle{1 \over 16\pi}\sin
\theta\{\left(-4r+M+2\lambda_0\right)+4r\}=\frac{M}{4}+\frac{\lambda_0}{2},\ee
which is exactly the energy of the first tetrad (22) as given by
Eq. (47).
 \newsection{Teleparallel Killing Vectors of the Bran-world  Spacetimes}
In this section we are going to calculate the Killing vectors of
the two tetrad  space-times, given by Eqs (22) and (23) to make
the picture more clear about the different results we obtained for
the energy and angular-momentum. Using the teleparallel Lie
derivatives of a covariant tensor of rank 2 established in
\cite{SJ} which is defined as \be \left({\it {{\cal L}^T}_\xi}
g\right)_{\mu \nu} \stackrel{\rm def.}{=}g_{\mu \nu ,\
\rho}\xi^\rho+g_{\rho \nu}{\xi^\rho}_{, \ \mu}+g_{\mu
\rho}{\xi^\rho}_{, \ \nu}+\xi^\rho\left(g_{\epsilon
\nu}{T^\epsilon}_{\mu \rho}+g_{\epsilon \mu}{T^\epsilon}_{\nu
\rho}\right),\ee where $,$ denote the ordinary derivative, $\xi$
is a vector field and ${T^\epsilon}_{\mu \rho}$ is the torsion
tensor defined by Eq. (7). The teleparallel Killing equations is
defined as \cite{SJ}\be \left({\it {{\cal L}^T}_\xi} g\right)_{\mu
\nu} \stackrel{\rm def.}{=}0.\ee Apply Eq. (63) to Eq. (22) we get
\ba 2 A'(r) \xi^1(r,\theta, \phi,t)+A(r) {\xi^0}_{,\
0}(r,\theta,\phi,t)\A=\A 0,  \nonu
A^2(r)B^2(r) {\xi}^0_{,\ 1}(r,\theta, \phi,t)- r{ \xi}^1_{,\
0}(r,\theta,\phi,t)-A(r)B^2(r)A'(r)\xi^0(r,\theta,\phi,t) \A=\A
0,\nonu
 A^2(r) {\xi^0}_{,\ 2}(r,\theta, \phi,t)-r^2 { \xi^2}_{, \
0}(r,\theta,\phi,t)\A=\A 0, \nonu
 A^2(r) {\xi^0}_{,\ 3}(r,\theta, \phi,t)-r^2 \sin^2 \theta
{\xi^3}_{, \ 0}(r,\theta, \phi,t)\A=\A 0,\nonu
(2rB'(r)-B(r)) \xi^1(r,\theta, \phi,t)-2rB(r){\xi^1}_{,\
1}(r,\theta,\phi,t)\A=\A 0,\nonu
B(r)\xi^2(r,\theta, \phi,t)(B(r)-\sqrt{r})+rB^2(r) {\xi^2}_{,\
1}(r,\theta, \phi,t)+{ \xi^1}_{,\ 2}(r,\theta,\phi,t) \A=\A 0,
\nonu
 B(r)\sin^2 \theta \xi^3(r,\theta, \phi,t)(B(r)-\sqrt{r})+
{\xi^1}_{,\ 3}(r,\theta, \phi,t)+rB^2(r)\sin^2\theta {\xi^3}_{,\
1}(r,\theta,\phi,t) \A=\A 0, \nonu
rB(r) {\xi^2}_{, \ 2}(r,\theta, \phi,t)+\sqrt{r} \xi^1(r,\theta,
\phi,t) \A=\A 0, \nonu
 {\xi^2}_{, \ 3}(r,\theta, \phi,t)+ \sin^2 \theta
{\xi^3}_{, \ 2} (r,\theta, \phi,t) \A=\A 0, \nonu
rB(r) \cos \theta \xi^2(r,\theta, \phi,t)+ rB(r) \sin \theta
{\xi^3}_{, \ 3}(r,\theta, \phi,t)+\sqrt{r}\sin \theta \xi^1
(r,\theta, \phi,t) \A=\A 0, \nonu
 \ea where  ${\xi^\alpha}_{, \ a}=\frac{\partial \xi^\alpha}
 {\partial x^a} $. From the first and fifth equations of  Eq. (64) we get
 \be \xi^1(r,\theta, \phi,t)=\frac{B(r)F(\theta,\phi,t)}{\sqrt{r}}, \qquad \xi^0(r,\theta,
 \phi,t)=-\int \frac{ 2A'(r)B(r)F(\theta,\phi,t)}{A(r) \sqrt{r}}dt+F_1(r,\theta,\phi),\ee
 using Eq. (65) in the sixth and seventh equations of Eq. (64) we get
\ba \xi^2(r,\theta, \phi,t)\A=\A
e^{\int{\left(\frac{\sqrt{r}-B(r)}{rB(r)}\right)dr}}\left\{-\int{\left(\frac{F_{,\
2}(\theta,\phi,t)e^{\int{\left(\frac{\sqrt{r}-B(r)}{rB(r)}\right)dr}
}}{r^{3/2}B(r)} \right)}dr+F_2(\theta,\phi,t)\right\}, \nonu
\xi^3(r,\theta, \phi,t)\A=\A
e^{\int{\left(\frac{\sqrt{r}-B(r)}{rB(r)}\right)dr}}\left\{-2\int{\left(\frac{F_{,\
3}(\theta,\phi,t)e^{\int{\left(-\frac{\sqrt{r}-B(r)}{rB(r)}\right)dr}
}}{r^{3/2}B(r)(\cos 2\theta-1)}
\right)}dr+F_3(\theta,\phi,t)\right\},
 \ea where $F(\theta,\phi,t)$, $F_1(r,\theta,\phi)$, $F_2(\theta,\phi,t)$ and
  $F_3(\theta,\phi,t)$ are arbitrary functions to be determined from the remaining
  equations of Eq. (64).
 Using Eqs. (65) and (66) in the remaining equations  of Eq. (64) we finally get
 \ba \xi^0(r,\theta, \phi,t)\A=\A C_0A(r), \quad \xi^1(r,\theta,
 \phi,t)=0, \quad
 \xi^2(r,\theta,\phi,t)=
\frac{{e^{\int \!{\frac {1}{\sqrt {r}B \left( r \right)
}}{dr}}}}{r}\left(C_1\sin\phi+C_2\cos\phi\right),\nonu
 \xi^3(r,\theta, \phi,t)\A=\A
 \frac{{e^{\int \!{\frac {1}{\sqrt {r}B \left( r \right)
}}{dr}}}}{r\sin\theta}\left(C_1 \cos \theta \cos\phi
 -C_2\cos\theta \sin\phi+C_3\sin\theta \right),\nonu
 \A \A \ea where $C_0, \
 C_1, \ C_2,$ and $C_3$ are four constants of integration. Thus \ba \xi \A=\A \Biggl(
C_0A(r)\frac{\partial}{\partial t}
 +\frac{{e^{\int \!{\frac {1}{\sqrt {r}B \left( r \right)
}}{dr}}}}{r}\left\{C_1\sin \phi+C_2\cos\phi\right\}
 \frac{\partial}{\partial \theta}\nonu
 \A \A +\frac{{e^{\int \!{\frac {1}{\sqrt {r}B \left( r \right)
}}{dr}}}}{r\sin\theta}\left\{C_1 \cos \theta \cos\phi
 -C_2\cos\theta \sin\phi+C_3\sin\theta \right\} \frac{\partial}{\partial
 \phi}\Biggr).\ea
 Eq. (68) gives the 4 Killing vector of the  spacetime given by
 Eq. (22) in the context of the TEGR as
 \ba \xi_{(1)} \A=\A A(r)\frac{\partial}{\partial t}, \nonu
\xi_{(2)} \A=\A \frac{{e^{\int \!{\frac {1}{\sqrt {r}B \left( r
\right) }}{dr}}}}{r\sin \theta }\left(\sin\theta \cos\phi
\frac{\partial}{\partial
 \theta}- \cos \theta \sin\phi  \frac{\partial}{\partial
 \phi}\right),\nonu
\xi_{(3)} \A=\A \frac{{e^{\int \!{\frac {1}{\sqrt {r}B \left( r
\right) }}{dr}}}}{r\sin \theta}\left(\sin\theta
\sin\phi\frac{\partial}{\partial
 \theta}+ \cos \theta \cos\phi  \frac{\partial}{\partial
 \phi}\right),\nonu
\xi_{(4)} \A=\A \frac{{e^{\int \!{\frac {1}{\sqrt {r}B \left( r
\right) }}{dr}}}}{r}\frac{\partial}{\partial \phi}. \ea

Now turn our attention to the derivation of the killing vector of
the second tetrad given by Eq. (23). Apply  Eq. (63) to Eq. (23)
 we get   \ba
2 A'(r) \xi^1(r,\theta, \phi,t)+A(r) {\xi^0}_{,\
0}(r,\theta,\phi,t) \A =\A 0,\nonu
 A^2(r)  B^2(r) {\xi}^0_{,\ 1}(r,\theta, \phi,t)- r{ \xi}^1_{,\
0}(r,\theta,\phi,t)-A(r) B^2(r)A'(r)\xi^0(r,\theta,\phi,t) \A =\A
0,\nonu
 A^2(R) {\xi^0}_{,\ 2}(r,\theta, \phi,t)-r^2 { \xi^2}_{, \
0}(r,\theta,\phi,t)\A =\A 0, \nonu
 A^2(r) {\xi^0}_{,\ 3}(r,\theta, \phi,t)-r^2 \sin^2 \theta
{\xi^3}_{, \ 0}(r,\theta, \phi,t)\A =\A 0,\nonu
\left(2rB'(r)-B(r)\right) \xi^1(r,\theta,
\phi,t)-2rB(r){\xi^1}_{,\ 1}(r,\theta,\phi,t) \A=\A 0,\nonu
B^2(r)\xi^2(r,\theta, \phi,t)+rB^2(r) {\xi^2}_{,\ 1}(r,\theta,
\phi,t)+{ \xi^1}_{,\ 2}(r,\theta,\phi,t) \A=\A 0, \nonu
B^2(r)\sin^2 \theta \xi^3(r,\theta, \phi,t)+ {\xi^1}_{,\
3}(r,\theta, \phi,t)+rB^2(r)\sin^2\theta {\xi^3}_{,\
1}(r,\theta,\phi,t) \A=\A 0, \nonu
 {\xi^2}_{, \ 2}(r,\theta, \phi,t) \A=\A 0, \nonu
{\xi^2}_{, \ 3}(r,\theta, \phi,t)- \sin^2 \theta {\xi^3}_{, \ 2}
(r,\theta, \phi,t)+{\xi^3} (r,\theta, \phi,t) \sin\theta \cos
\theta  \A=\A 0, \nonu
  {\xi^3}_{, \ 3}(r,\theta, \phi,t) \A=\A 0.  \ea
 From the first, fifth, eighth and tenth equations  of  Eq. (70) we get
 \ba \xi^1(r,\theta, \phi,t)\A=\A \frac{B(r)}{\sqrt{r}}F_4(\theta,\phi,t), \qquad \xi^0(r,\theta,
 \phi,t)=-\int \frac{2A'(r)B(r)}{\sqrt{r}A(r)}F_4(\theta,\phi,t)dt+F_5(r,\theta,\phi), \nonu
 \xi^2(r,\theta, \phi,t)\A=\A F_6(r,\phi,t), \qquad \xi^3(r,\theta, \phi,t)=
 F_7(r,\theta,t),
 \ea  where $F_4(\theta,\phi,t)$, $F_5(r,\theta,\phi)$,
 $F_6(r,\phi,t)$ and $F_7(r,\theta,t)$ are arbitrary
 functions to be determined from the remaining equations of Eq. (70).
 Using Eq. (71) in the remaining equations of Eq. (70) we finally get

 \be \xi^0(r,\theta, \phi,t)=0, \qquad \xi^1(r,\theta,
 \phi,t)=0, \qquad
   \xi^2(r,\theta,\phi,t)= \frac{C_4}{r}, \qquad
 \xi^3(r,\theta, \phi,t)=\frac{C_5}{r\sin\theta},\ee
 where $C_4$ and $C_5$ are two constants of integration. Thus
\ba \xi \A=\A \Biggl(\frac{C_4}{r} \frac{\partial}{\partial
\theta} +\frac{C_5}{r\sin\theta}\frac{\partial}{\partial
 \phi}\Biggr).\ea

 Eq. (73) gives the 2 Killing vector of the space-time given by
 Eq. (23) in the context of the TEGR as
 \be \xi_{(1)} = \frac{1}{r}\frac{\partial}{\partial
 \theta},\qquad \xi_{(2)}= \frac{1}{r\sin \theta } \frac{\partial}{\partial
 \phi}. \ee
 It is well known that Minkowski spacetime has the Poincar$\acute{e}$
 symmetry algebra with the 10 generators
 \cite{KMQ,Hg}
 \ba \xi_0 \A=\A \frac{\partial}{\partial t}, \qquad \xi_1=\cos \phi
 \frac{\partial}{\partial \theta}-\cot \theta \sin \phi\frac{\partial}{\partial
 \phi}, \nonu
\xi_2 \A=\A\sin \phi
 \frac{\partial}{\partial \theta}+\cot \theta \cos \phi\frac{\partial}{\partial
 \phi}, \qquad \xi_3=\frac{\partial}{\partial \phi}, \nonu
 \xi_4 \A=\A \sin \theta \cos \phi
 \frac{\partial}{\partial r}+\frac{\cos \theta \cos \phi}{r}\frac{\partial}{\partial
 \theta}-\frac{\csc \theta \sin \phi}{r}\frac{\partial}{\partial
 \phi},\nonu
 \xi_5 \A=\A \sin \theta \sin \phi
 \frac{\partial}{\partial r}+\frac{\cos \theta \sin \phi}{r}\frac{\partial}{\partial
 \theta}+\frac{\csc \theta \cos \phi}{r}\frac{\partial}{\partial
 \phi},\nonu
\xi_6 \A=\A \cos \theta
 \frac{\partial}{\partial r}-\frac{\sin \theta }{r}\frac{\partial}{\partial
 \theta},\nonu
\xi_7 \A=\A r \sin \theta \cos \phi
 \frac{\partial}{\partial t}+t\left\{\sin \theta \cos \phi \frac{\partial}{\partial
 r}+\frac{\cos \theta \cos \phi}{r}\frac{\partial}{\partial
 \theta}-\frac{\csc \theta \sin \phi}{r}\frac{\partial}{\partial
 \phi} \right\},\nonu
\xi_8 \A=\A r \sin \theta \sin \phi
 \frac{\partial}{\partial t}+t\left\{\sin \theta \sin \phi \frac{\partial}{\partial
 r}+\frac{\cos \theta \sin \phi}{r}\frac{\partial}{\partial
 \theta}+\frac{\csc \theta \cos \phi}{r}\frac{\partial}{\partial
 \phi} \right\},\nonu
 \xi_9 \A=\A r \cos \theta
 \frac{\partial}{\partial t}+t\left\{\cos \theta \frac{\partial}{\partial
 r}-\frac{\sin \theta}{r}\frac{\partial}{\partial
 \theta} \right\},
 \ea
 where the speed of light $c=1$ and  $\xi_0, \xi_4, \xi_5$ and  $\xi_6$ are the spacetime
 translations which provide the laws of conservation of energy and
 linear momentum, $\xi_1, \xi_2$ and $\xi_3$ are the rotations
 which provide the laws of conservation of angular momentum and
 $\xi_7, \xi_8$ and $\xi_9$ are the Lorentz transformations which
 provide the laws of conservation of spin and angular momentum via
 Noether's theorem \cite{FMQ}. {\it For a spherically symmetric space-time only
 the 4 generators $\xi_0, \xi_1, \xi_2$ and $\xi_3$ apply, yielding
 only conservation of energy and angular momentum. The generator
 $\xi_4, \xi_5,  \xi_6, \xi_7, \xi_8$ and $\xi_9$ yielding
 conservation of linear momentum and spin angular momentum
  are lost for this space-time.}

  Let us compare our results with the first four equations of Eq.
  (75). For the tetrad field given by Eq. (22)  we get
  the 4 Killing vectors to have the form given by Eq. (69). This
  Killing vectors are in consistence with the first 4 Killing
  vectors of Eq. (75) when $A(r)=1$ and $B(r)=\sqrt{r}$. Energy,
   irreducible mass,  momentum and angular momentum associated
   with the tetrad given by Eq. (22) are satisfactory and in consistence with
  the previous results \cite{SNH}. So this tetrad has no problem either for
  the calculations of energy, irreducible mass, momentum and angular momentum
  or for the associated Killing vectors which given by Eq. (69).

  The tetrad given by Eq. (23) has many problems: First the energy is not equal
  the ADM. Also the irreducible mass is effected by the horizon and some
   components of the angular momentum are not
  vanishing. This may be due to the fact that the Killing
  vector associated with tetrad (23) is not in
  agreement with that given by Eq. (75). Since the first four equations of Eq. (75)
  are the responsible for the conservation of energy and angular momentum and
  these equations are disappear from  tetrad (23) as is clear from by Eq. (74).

\newsection{Main results and Discussion}
The main results of this paper can be summarized as follow:\vspace{0.4cm}\\
$\bullet$  Two different tetrad fields are used. These tetrads
are related by a local Lorentz transformation which keeps
spherical symmetry, i.e., the tetrad (22) can be written in terms
of the tetrad (23) using the following local Lorentz
transformation \be \left({{e_1}_a}^{ \mu} \right) =
{\Lambda_\nu}^\mu \left({{e_2}_a}^{ \nu} \right), \ where \
{\Lambda_\nu}^\mu=\left( \matrix{ 1&0 & 0 & 0 \vspace{3mm} \cr 0 &
\sin\theta \cos\phi & \cos\theta \cos\phi
 & -\sin\phi \vspace{3mm} \cr
0 &  \sin\theta \sin\phi & \cos\theta \sin\phi
 & \cos\phi \vspace{3mm} \cr
0 &  \cos\theta & -\sin\theta & 0 \cr } \right). \ee
 The space-time
associated with these
tetrad fields is  given by Eq. (24). \vspace{0.3cm}\\
$\bullet$  The energy of these tetrad fields are calculated using
the gravitational energy-momentum tensor, which is a coordinate
independent \cite{Mj}. One of this tetrad field given by Eq. (22)
gives a satisfactory results for the energy, while the other which
is given by Eq. (23) its associated energy depends on the radial
coordinate.
\vspace{0.3cm}\\
$\bullet$ Calculations of the torsion components associated with
these two tetrad fields  are given. From these calculations we
show that the torsion components of each tetrad field are
different. This may gave an indication why the energy of the two
tetrad fields is different.
 \vspace{0.3cm}\\
 $\bullet$ We use the regularized expression of the gravitational energy-momentum
 tensor to calculate the energy associated with the second tetrad field given by Eq. (23).
 \vspace{0.3cm}\\
$\bullet$ After using the regularized expression of the
gravitational energy-momentum tensor we show that the energy
associated with the two tetrad fields is equal. \vspace{0.3cm}\\
$\bullet$ Using the definition of the energy and the angular
momentum given by Eqs. (13) and (20) we show by explicit
calculations that the angular momentum  depends on the choice of
the frame used. \vspace{0.3cm}\\
$\bullet$ The calculations of the irreducible mass is given within
the external horizons.  From this calculations we show that the
external horizons  do not play any role on the energy
\vspace{0.2cm}\\
$\bullet$ To make the picture more clear we have calculated the
Killing vectors associated with these tetrad fields using  the
definition of the Lie derivative of a second rank tensor in the
framework of TEGR \cite{SJ}.
 We have shown by explicit calculations that the tetrad which show consistence results
 of energy, irreducible mass, momentum and angular momentum continue to
 show consistency for the Killing vector with GR \cite{KMQ}. The other tetrad fields
 also continue to give results of Killing vectors in contradiction  with
 that of GR.  This contradiction clear why the results of energy, irreducible mass,
 momentum and angular momentum are not in consistence. \vspace{0.3cm}\\


\newpage
\begin{thebibliography}{99}
\bibitem{KK} T. Kaluza, {\it Sitzungseber. Press. Akad. Wiss.
Phys. Math. Klasse K1} (1920), 966; O. Klein, {\i Z. Phys.} {\bf
37} (1926), 895.

\bibitem{HW} P. Ho$\tilde{r}$ava and E. Witten, {\it Nucl.
Phys.} {\bf B460}, (1996), 506; {\bf B475} (1996) 94.

\bibitem{RR} L. Randall and R. Sundrum, {\it Phys.  Rev. Lett. }
{\bf 83} (1999) 3370; 4690.

\bibitem{SMS} T. Shiromizu, K. Maeda and M. Sasaki, {\it Phys.
Rev. }{D62} (2000), 024012.

\bibitem{GC} K.A. Bronnikov, V.N. Melnikov and H. Dehnen, {\it
Phys. Rev. } {\bf D68} (2003), 024025.

\bibitem{BGCF} K.A. Bronnikov, G. Cl$\acute{e}$ment, C.P.
Constantinidis and J.C. Favris, {\it Phys. Lett.}{A243} (1998),
121; {\it Grav. Cos.} {\bf 4} (1998), 128.


\bibitem{Kt} T.W.B. Kibble, {\it J.Math. Phys.} {\bf 2} (1961)
212.
\bibitem{HHKN} F.W. Hehl, P. Von der Heyde, D. Kerlick and J.
Nester, {\it Rev. Mod. Phys.} {\bf 48} (1976), 393.

\bibitem{Hf8} F.W. Hehl, in: {\it General Relativity  and Gravitation- One Hundred Years
after the birth of Albert Einstein, ed. } A. Held (Plenum, New
York, 1980) Vol 1.

\bibitem{HS} K. Hayashi, and T. Shirafuji, {\it Prog.\ Theor.\ Phys.\ }{\bf 64} (1980),
866, 883, 1435, 2222; {\bf 65}, 525.

\bibitem{BN} M. Blagojevi$\acute{c}$  and I.A. Nikoli$\acute{c}$   {\it Phys.
Rev. } {\bf D62} (2000), 024021.

\bibitem{HNV} F.W. Hehl, J. Nitsch and  P. von der Heyde,
in {\it General Relativity and Gravitation}, A.\ Held, ed. (Plenum
Press, New York) (1980).

\bibitem{HMM} F.W. Hehl, J.D. MacCrea, E.W. Mielke and Y. Ne'eman, {\it Phys. Rep.}
{\bf 258} (1995), 1.

\bibitem{HS9} K. Hayashi and  T. Shirafuji,  {\it Phys.\ Rev.\ } {\bf D19} (1979), 3524.

\bibitem{Nj} J. Nitsch, in: {\it Cosmology and Gravitation: Spin,
Torsion, Rotation and Supergravity, eds. } Bergman and V. de
Sabbata (Plenum, New York, 1980).

\bibitem{BT} P.G. Bergmann and R. Thomson, {\it Phys. Rev.} {\bf 89} (1953),
401.
\bibitem{LL} L.D. Landau and E.M. Lifshitz, {\rm The Classical Theory of
Fields (Pergamon Press, Oxford, 1980)}.

\bibitem{PP}  C. Pellegrini and J. Plebanski, {\it Mat.\ Fys.\
 Scr.\ Dan.\ Vid.\ Selsk.\ }{\bf 2} (1963), no.3.

\bibitem{Mo} C. M\o ller, (1961) {\it Ann. of Phys. } {\bf 12} (1961), 118.

\bibitem{Mo2}  C. M\o ller,  {\it ``Tetrad fields and conservation laws in
general relativity"} in Proc. International School of Physics
``Enrico Fermi" ed. C. M\o ller, (Academic Press, London, 1962).

\bibitem{Mo3} C. M\o ller, {\it Mat.\ Fys.\ Medd.\ Dan.\ Vid.\ Selsk.\ } {\bf 1} (1961), 10.

\bibitem{Mo4} C. M\o ller, {\it Nucl.\ Phys. } {\bf 57} (1964), 330.

\bibitem{HN} K. Hayashi and  T. Nakano, {\it Prog.\ Theor.\ Phys.\ }{\bf 38}
(1967), 491.

\bibitem{Kw} W. Kopzy$\acute{n}$ski, {\it J.\ Phys.\ } {\bf A15} (1982),
493.

\bibitem{NHC} C.C. Chang, J.M. Nester and C.M. Chen, {\it
Phys. Rev. Lett.} {\bf 83} (1999), 1897;  R.S. Tung and J.M.
Nester, {\it Phys. Rev. } {\bf D60} (1999) 021501; J.M. Nester and
H.J. Yo Chin, {\it J. Phys.} {\bf 37} (1999) 113; J.M. Nester, F.
H. Ho and C. M.Chen, {\it Quasilocal Center-of-Mass for
Teleparallel Gravity,  Proceeding of the 10th. Marcel Grossman
Meeting (Rio de Janeiro, 2003)} gr-qc/0403101; J. M. Nester, {\it
Phys. Lett.} {\bf A139} (1989) 112; {\it J. Math. Phys.} {\bf 33}
(1992), 910; {\it Class.\ Quantum Grav.\ } {\bf 5} (1988), 1003..

\bibitem{Tn} N. Toma,  {\it Prog.\ Theor.\ Phys.\ }
{\bf 86} (1991), 659; T. Kawai and N. Toma,  {\it Prog.\ Theor.\
Phys.\ } {\bf 87} (1992), 583 .

\bibitem{AGP1} V.C. de Andrade and J.G. Pereira, {\it Phys.\ Rev.\ } {\bf D56}
(1997), 4689; V.C. de Andrade, L.C.T Guillen  and J.G. Pereira,
{\it Phys.\ Rev.\ Lett.\ } {\bf 84} (2000), 4533;  {\it Phys.\
Rev.\ } {\bf D64} (2001), 027502.

\bibitem{MR} J.W. Maluf and J.F. da Rocha-Neto, {\it Phy. Rev.} {\bf D64}
(2001), 084014.

\bibitem{Sj} J. Schwinger, {\it Phys. Rev.} {\bf 130} (1963),
1253.

\bibitem{Mj} J. W. Maluf, {\it J.\ Math.\ Phys.\ }{\bf 35} (1994), 335;
J. W. Maluf and A. Kneip, {\it J.\ Math.\ Phys.\ }{\bf 38} (1997),
458; J. W. Maluf and J. F. da Rocha-Neto, {\it J. Math. Phys.
}{\bf 40} (1999), 1490; J. W. Maluf and A. Goya,  {\it Class.
Quant. Grav. }{\bf 18} (2001), 5143; J. W. Maluf, J. F. da
Rocha-Neto, T. M. L. Toribio and K. H. Castello-Branco,  {\it
Phys.\ Rev.\ }{\bf D65} (2002), 124001;  A. A. Sousa and J. W.
Maluf, {\it Prog.\ Theor.\ Phys. }{\bf 108} (2002), 457.

\bibitem{RC} J.F.  da Rocha-Neto and K. H.
Castello-Branco,  {\it JHEP }{\bf 0311} (2003), 002.

\bibitem{RT} T. Regge and C. Teitelboim, {\it Ann. Phys. (New
York)} {\bf 88} (1974), 286.

\bibitem{Yoj} J. W. York, {\it Jr., Energy and Momentum of the
Gravitational field, in ``Essays in General Relativity"} edited by
F.J. Tipler (Acdemic Press, New York, 1980).

\bibitem{BM} R. Beig and N. $\acute{o}$ Murchadha, {\it Ann. Phys.
(New York)} {\bf 174} (1987), 463.

\bibitem{SLB} L.B. Szabados, {\it Class. Quantum. Grav.} {\bf 20}
(2003) 2627.

\bibitem{MS2} J. W. Maluf and A.A. Sousa, {\it Hamiltonian formulation of teleparallel
theories of gravity in the time gauge} gr-qc/0002060, (2000).

\bibitem{SM2}  A. A. Sousa and J. W. Maluf, {\it Prog.\ Theor.\ Phys. }{\bf 104} (2000), 531.

\bibitem{MD} J. W. Maluf and J. F.  de Rocha-Neto, {\it Phys.\ Rev.\ }{\bf D64} (2001), 084014.

\bibitem{MUFR} J.W. Maluf, S.C. Ulhoa, F.F. Faria and J.F. da Rocha-Neto,
{\it Calss. Quant. Grav.} {\bf 23} (2006), 6245.

\bibitem{MPR} A. A. Sousa, R.B. Pereira  and J.F. da Rocha-Neto, {\it Prog. Theor. Phys. }
{\bf 114} (2005), 1179; A. A. Sousa, J.S. Moura and R.B. Pereira
gr-qc/0702109.

\bibitem{SJr} Shanxian Xu and Jiliang Jing, {it Class.\  Quant.\ Grav.} {\bf
23} (2006), 4659.

\bibitem{Nj8} J. M. Nester, {\it Int. J. Mod. Phys. }{\bf A4} (1989),
1755.

\bibitem{BV2001} M. Blagojevi$\acute{c}$  and M. Vasili$\acute{c}$ {\it Phys. Rev.
} {\bf D64} (2001), 044010.

\bibitem{CN2} C.C. Change, J.M. Nester, {\it Grav. $\&$ Cosmol.}
{\bf 6} (2000) 257.

\bibitem{SNC} L.L. So, J.M. Nester and H. Chen, {\it The 7th.
conference on Gravitation and Astrophysics; gr-qc/0605150.}

\bibitem{SL} L.B. Szabados, ``Quasi-Local Energy-Momentum and Angular Momentum in GR: A Review Article",
Living Rev. Relativity 7,  (2004), 4.
http://www.livingreviews.org/lrr-2004-4.


\bibitem{Mj5} J. W. Maluf, {\it Annalen Phys.} {\bf 14} (2005),
723; {\it Gravitational and Cosmology} {\bf 11} (2005), 284.

\bibitem{CD} D. Christodoulou, {\it Phys. Rev. Lett. } {\bf 25}
(1970), 1596.

\bibitem{BG} G. Bergqvist, {\it Class. Quant. Grav.} {\bf 9}
(1992), 1753.

\bibitem{Ka} A. Komar, {\it Phys. Rev.} {\bf 113} (1959), 934; J.
Winicour, ``$\,$ Angular Momentum in General Relativity", in
General Relativity and Gravitation edited by A. Held (Plenum, New
York, 1980); A. Ashtekar, ``$\,$ Angular Momentum of Isolated
Systems in General Relativity", in Cosmology and Gravitation,
edited by P.G. Bergmann and V. de Sabbata (Plenum, New York 1980).

\bibitem{HS8} K. Hayashi and T. Shirafuji, {\it Prog. Theor.
Phys.} {\bf 73} (1985), 54;  M. Blagojevi$\acute{c}$ and M.
Vasili$\acute{c}$  {\it Class. Quant. Grav.} {\bf 5} (1988), 1241;
T. Kawai, {\it Phys. Rev. } {\bf D62} (2000), 104014, T. Kawai, K.
Shibata and I. Tanaka, {\it Prog. Theor. Phys.} {\bf 104} (2000)
505.

\bibitem{Dp} P.A.M. Dirac,  {\it Lectures on Quantum Mechanics (Belfer Gradute School
of Science (Monographs Series No. 2) Yeshiva University, New York,
1964) }.

\bibitem{Rh} H.P. Robertson,  {\it Ann.\ of Math.\ (Princeton)} {\bf 33}
(1932), 496.

\bibitem{HS7} K. Hayashi and  T.  Shirafuji, {\it Phys.\ Rev.\ }{\bf D19}
(1979), 3524.

\bibitem{SNH} T. Shirafuji, G.G.L. Nashed, and K. Hayashi,   {\it Prog.\
Theor.\ Phys.\ } {\bf 95} (1996), 665.

\bibitem{SA} M. Salti and O. Aydo$\hat{g}$du M. Korunur, {\it  will appear in JHEP};
gr-qc/0611014.

\bibitem{ADM} R Arnowitt, S. Deser and C.W. Misner, {\it
Gravitation: An Introduction to Current Research} edited by L.
Witten (Wiley, N.Y. (1962)).

\bibitem{MTW} C.W. Misner, K.S. Thorne and J.A. Wheeler, {\it
Gravitation (Freeman, San Francisco, 1973)}, P. 435.

\bibitem{CFM} R. Casadio, A. Fabbri and L. Mazzacurati, {\it Phys.
Rev. } {\bf D65} (2002), 084040.

\bibitem{GM} C. Germani and R. Maartens, {\it Phys. Rev. }{\bf
D64} (2001), 124010.

\bibitem{BK} K.A. Bronnikov and S.-W. Kim, {\it Phys. Rev. }{\bf
D67} (2003), 064027.

\bibitem{MVR}  J. W. Maluf, M.V.O. Veiga and J. F. da Rocha-neto, {\it Gen. Rel.
Grav.} {\bf 39} (2007), 227.

\bibitem{BHS} P. Bae(c)kler, R. Hecht, F.W. Hehl and T. Shirafuji,
{\it Prog. Theor. Phys.} {\bf 78} (1987), 16.

\bibitem{YB} J.D. Brown and J.W. York, Jr., {\it Quasi-local
energy in general relativity}, Proceedings of the Joint Summer
Research conference on Mathematical Aspects of Classical Field
Theory, edited by M.J. Gotay, J.E. Marsden and V. Moncrief
(American Mathematical Socity, 1991); {\it Phys. Rev. } {\bf D47}
(1993), 1407.

\bibitem{SJ} M. Sharif and M. Jamil Amir, {\it Mod. Phys. Lett.
A} {\bf 23} (2008), 969.

\bibitem{KMQ} D. Kramer, H. Stephani, M.A.H. MacCullum and E.
Herlt, {\it Exact solutions of Einstein field equations, Cambridge
Univ. Press, Cambridge, 1980}.

\bibitem{Hg} G.S. Hall, {\it Symmetryies and curvature structure in
general relativity, World Scientific , Singapore, 2004}.

\bibitem{FMQ} T. Feroze, F.M. Mahomed and A. Qadir, {\it Nonlinear
Dynam. }{\bf 45} (2006), 65.


\bibitem{Mo66} C.  M\o ller,
 {\it Mat.\ Fys.\ Medd.\ Dan.\ Vid.\ Selsk.\ }{\bf 35} (1966), no.3.

\end{thebibliography}
\end{document}